\documentstyle{elsart}
\begin{document}

\begin{frontmatter}

\title{SLOW FORCING IN THE PROJECTIVE DYNAMICS METHOD}

\author[SCRI]{M.~A.\ Novotny}
\author[SLOV]{M.~Kolesik} and
\author[SCRI,MARTECH]{P.~A.\ Rikvold}

\address[SCRI]{Supercomputer Computations Research Institute (SCRI),
Florida State University, Tallahassee, FL 32306-4130, U.S.A.}
\address[SLOV]{Institute of Physics, Slovak Academy of Sciences, 
D\'ubravsk\'a cesta 9,
84228 Bratislava, Slovak Republic}
\address[MARTECH]{Center for Materials Research and Technology (MARTECH) 
and Department of Physics,
Florida State University, Tallahassee, FL 32306-4350, U.S.A.}

\begin{abstract}
We provide a proof that when there is no forcing the 
recently introduced projective 
dynamics Monte Carlo algorithm gives the exact lifetime of 
the metastable state, within statistical uncertainties.  
We also show numerical evidence illustrating that for slow forcing 
the approach to the zero-forcing limit 
is rather rapid.  The model studied numerically is the 
3-dimensional 3-state Potts ferromagnet.  
\end{abstract}


\end{frontmatter}

\vfill
\eject

One of the most challenging hurdles to be overcome 
in computational materials science is finding a direct way to bridge 
the disparate time and length scales required in 
simulations of models of materials.  The length 
scales go from the atomic to the size of the actual material sample.  
The shortest time scale in many instances can be associated 
with an inverse phonon frequency, and 
the longest time scale with the mean time before macroscopic material failure.  
One way of bridging these time and length scales is to devise 
better algorithms.  Such algorithms need to increase 
computational efficiency by orders of magnitude in order to 
allow the desired material simulations to proceed on today's computers.  

Although there are a number of efficient algorithms for the 
{\it statics\/} of model 
systems \cite{SWF1}, the number of advanced algorithms for dynamics is 
more limited.  These include
the $n$-fold way algorithm \cite{NFOLD}, and its generalization, 
the Monte Carlo with Absorbing Markov Chains (MCAMC) algorithm \cite{MCAMC}.  

Recently, the authors have devised an improved dynamical algorithm 
for models with discrete numbers of states at each lattice site.  
This algorithm is called the Projective Dynamics (PD) method \cite{MIROPRL}.  
The PD algorithm builds on previous work using projections onto 
slow variables in dynamical models \cite{SCHUL,LEE}.  
The PD algorithm has been shown to allow simulations of 
the dynamics of systems such as the Ising model 
for very long times and large lattice sizes \cite{MIROUGA}, 
requiring quite modest computer resources.  
It has also been shown to allow identification of 
physical quantities, such as activation volumes and Barkhausen volumes, 
in dynamical models of magnetic domain growth \cite{MIROMRS}.  

In this brief paper we concentrate on the use of the PD algorithm 
to obtain the average lifetime of the metastable state in a model system.  
In particular, we apply the PD algorithm to the escape from the 
metastable state of the three-dimensional 3-state Potts ferromagnet.  

The 3-state Potts model in applied field $H$ has the Hamiltonian
\begin{equation}
{\cal H} = 
-J \sum_{\langle i,j \rangle} \delta (\sigma_i,\sigma_j)
+H \sum_i\left[\delta(0,\sigma_i)-\delta(1,\sigma_i)\right]
\label{eq:HAM}
\end{equation} 
where $\sigma_i \in \{0,1,2\}$ is the `spin' at
lattice site $i$.  The first summation runs over all
nearest-neighbor pairs on a simple-cubic lattice, 
and $J$$>$$0$ corresponds to ferromagnetic interactions.  
The second sum runs over all lattice sites, and 
it only changes the energies of spins in the states 0 and 1.  
The system can be characterized by the
concentrations $\{n_0,n_1,n_2\}$, $\sum n_i$$=$$1$, of
spins in the three states.  

The Monte Carlo simulation uses 
local updates at randomly chosen sites.  
We use simple cubic lattices of size $L^3$$=$$V$ with periodic 
boundary conditions.  
In this paper we use Glauber dynamics, and we measure the 
time in Monte Carlo Steps per Spin (MCSS).  
The simulation starts with all spins in state 0, 
with the temperature $T$ below the critical temperature $T_{\rm c}$.  
We measure the 
average lifetime $\tau$, defined as the average 
first-passage time to a configuration 
with half of the spins in 
the stable state, which here is state 1.  
Our Monte Carlo simulation 
can be viewed as a random walk which starts in the state 
with $n_0$$=$$V$ 
and goes until it reaches a state with $n$$=$$N$$=$$V/2$, where 
we have dropped the subscript on $n$$=$$n_1$.  

The spins in the current configuration can be divided into classes
specified 
by the state $\sigma$ of the spin 
and by the numbers 
$\{a,b,6-a-b\}$ of neighbors in the states $\{0,1,2\}$.  
Let $c^{\sigma}_{a b}(n_0,n_1,n_2)$ be
the average concentration of spins in the
class $\{\sigma,a,b\}$, conditional on the total
concentrations $n_i$ of spins in state $i$. 
This average is obtained from repeated metastable simulations.  
The class concentrations are normalized such that 
$\sum_{\sigma a b}c^\sigma_{ab}(n_0,n_1,n_2)$$=$$1$ for all 
$\left\{n_0, n_1, n_2\right\}$.  
Let $p^{\sigma \sigma'}_{a b}$ be the probability
that a spin in the class $\{\sigma,a,b\}$ will flip to
the state $\sigma'$ when visited by the updating
algorithm.
We project the spin class concentrations onto bin $n$ by defining 
$c^{\sigma}_{ab}(n)$$=$$\sum_{n_0,n_2}c^{\sigma}_{ab}(n_0,n_1$$=$$n,n_2)$.  
The projected flip rates are defined 
in terms of the 
$c^{\sigma}_{ab}(n)$ 
and the flipping probabilities
$p^{\sigma \sigma'}_{ab}$ as: 
\begin{equation}
g(n)=\sum_{ab,\sigma\ne 1} c^{\sigma}_{ab}(n)\  p^{\sigma 1}_{ab}\ , 
\quad \quad 
s(n)=\sum_{ab,\sigma\ne 1} c^{1}_{ab}(n)\  p^{1 \sigma}_{ab} .
\label{eq:gsrates}
\end{equation}
The rates $s(n)$ and $g(n)$ correspond to the 
shrinkage and growth rates of the stable phase, respectively.  They depend 
on how the configurations are generated to obtain the 
$c^{\sigma}_{ab}(n)$.  Here we obtain the configurations 
during simulation of escape from the metastable state, 
with and without the forcing described below.  

Using standard methods from the theory of absorbing
Markov chains \cite{MCAMC,LEE}, the mean
lifetime $\tau$ and the total average time $h(n)$
(measured in MCSS, with $h(N)=0$) spent by the random
walker in the state $n$ are given, as shown in the appendix, by 
\begin{equation} 
\label{TAUETC}
 \tau = \sum_{n=0}^{N-1} h(n)\ \ , 
\quad \quad
 h(n-1) = { V^{-1} + s(n) h(n) \over g(n-1)} \ .  
\label{eq:lifetime}
\end{equation} 

An enhancement to the PD algorithm described above was 
presented in Ref.~\cite{MIROPRL}.  At any time, the system is only allowed 
to have the number of spins in the stable phase, $n$, 
strictly larger than 
a lower bound $n_{\rm min}(t)$.  The simulation starts with 
$n_{\rm min}$$=$$-1$, and $n_{\rm min}$ is increased 
at a slow constant rate.  The class concentrations 
$c^\sigma_{a b}(n)$ are measured during this forced escape and used to 
calculate $\tau$ from Eq.~(\ref{TAUETC}).  
In the appendix we prove that 
if the random walk does not hit the moving lower bound,  
the value of $\tau$ obtained with the PD algorithm is exact.  
Figure~1 shows the approach of $\tau$ to the 
slow-forcing limit as the forcing rates is lowered.  Even at 
moderately high forcing rates the lifetime (with 
statistics between $10^2$ and $10^3$ 
escapes from the metastable state) is 
indistinguishable from the value at zero forcing rate.  The forcing, 
however, decreases substantially the amount of computer time 
required to obtain $\tau$.  
In the conventional direct simulation approach, the necessary computer 
time increases {\it exponentially\/} with the depth of the metastable 
free-energy minimum.  On the other hand, in the forced-escape method, 
the computer time is proportional to the 
depth of the free-energy minimum itself.  
In other words, the proposed method allows calculations of an 
exponentially hard problem in linear time.  
Although here we only present data for a 
temperature and field at which direct simulation is possible and the 
speed-up is only about $10^2$--$10^3$, the forced escape 
method allows one to obtain lifetimes many orders of magnitude longer 
than would be possible without forcing \cite{MIROPRL,MIROUGA}.  

In conclusion, we have obtained the average lifetime, $\tau$, of the 
metastable state of the 3-state three-dimensional Potts ferromagnet 
using forcing in the PD (Projective Dynamics) method.  In the limit of 
zero forcing, we prove in the appendix that the value of $\tau$ obtained 
is exact.  
Substantial savings in computer time can 
be achieved using relatively moderate forcing rates without 
significantly affecting the obtained values of $\tau$.  

\vskip 15 truept
\noindent {\bf ACKNOWLEDGEMENTS}

This research was supported in part by U.S.\ NSF Grants 
9520325 and 9871455,
and by the Florida State University through MARTECH and through 
SCRI (DOE Contract No.\ DE-FC05-85ER25000).
MK was also supported by VEGA grant no.~2/4109/97.  
Supercomputer access provided by the DOE at NERSC.

\vskip 15 truept
\noindent {\bf APPENDIX:  ~ Lifetimes from projective dynamics}

We prove that if there is no forcing, the lifetime obtained 
from the projective dynamics algorithm is exact.  
Of course, there will still be statistical uncertainties 
in the lifetime obtained from any finite length simulation.  
The proof is given for the model under consideration in this paper, 
the three-dimensional three-state Potts model.  
Furthermore, we choose state~1 as the slow variable 
on which we project the dynamics.  
The proof 
can easily be generalized to other models and dimensions as long as 
the number of states in the model is finite.  
The notation we use is given below.  
\begin{itemize}
\item[$\bullet$] $V$  --- total number of spins in the system.  
\item[$\bullet$] $S = \bigcup_{n=0}^V S_n $ ---
partition of the state space $S$ of the model into disjoint subspaces
where $S_n$ contains all microstates with exactly $n$ spins in state 1.
\item[$\bullet$] $[n,i]$ --- 
label for the $i$th state in $S_n$, $i=1, 2, \ldots, |S_n|$, 
where $|\cdots|$ denotes the number of elements in the partition.  
\item[$\bullet$] 
 $C^{\sigma}_{ab}(n,i)$ --- the number of those spins of the microstate
 $[n,i]$ which are in state $\sigma$ and have $a$ and $b$ neighbors in states
 0 and 1, respectively. Thus, $\sum_{ab\sigma} C^{\sigma}_{ab}(n,i) = V$
 for each $[n,i]$.  
\item[$\bullet$]  $c^{\sigma}_{ab}(n,i) = C^{\sigma}_{ab}(n,i)/V$ 
denotes concentrations.
\item[$\bullet$]   $p_{ab}^{\sigma \sigma '}$ --- 
probability that a chosen spin
in state $\sigma$ having $a,b$ neighbors in states 0 and 1 will flip
to state $\sigma '$.
\end{itemize}

\noindent
Consider direct Monte Carlo measurement of the mean lifetime
$\tau$ without forcing.  This can be accomplished by setting 
$n_{\rm min}=-1$ at all times.  
Start an escape from the state $[0,1]$ and continue 
the simulation until the Monte Carlo 
``random walker'' reaches one of the absorbing states 
$[n_{\rm stop},j]$ with $j$$=$$1, 2, \ldots, |S_{n_{\rm stop}}|$.  
A total number $N_{\rm esc}$ of independent escapes are realized. 
Denote by $W(n,i)$ the number of visits in the 
state $[n,i]$ generated by this repeated measurement 
(counting rejected Monte Carlo moves as new visits in the same state).  
Then 
\begin{equation}
h(n) = {1\over V} \sum_{i=1}^{|S_n|} W(n,i)/N_{\rm esc}
\end{equation} 
is the average (sojourn) time spent in $S_n$, 
in units of Monte Carlo Steps per Spin (MCSS).  
The mean lifetime is obtained by 
\begin{equation}
\tau = \sum_{n=0}^{n_{\rm stop}-1} h(n) =
     {1\over V} \sum_{n=0}^{n_{\rm stop}-1} 
\sum_{i=1}^{|S_n|} W(n,i)/N_{\rm esc} \ .
\end{equation}
One can imagine that each random walker generates an oriented ``world line''
starting in $[0,1]$ and ending in some $[n_{\rm stop},j]$, because the
transitions are possible only between neighboring $S_n$ and 
$S_{n\pm 1}$ subspaces. 
Therefore, an ``equation of continuity'' holds for each $n<n_{\rm stop}$:
\begin{equation}
\label{NNP1}
N_{n\to n+1} = N_{\rm esc} + N_{n+1 \to n}  
\end{equation}
where $N_{\ell\to \ell'}$ stands for the number of transitions between the
subspaces with $\ell$ and $\ell'$ spins in state~1. 
Equation~({\ref{NNP1}}) can be written in terms of 
$W$, $C_{ab}^{\sigma}$ and $p_{ab}^{\sigma \sigma '}$.  
To do this we observe that the probability that a particular attempted spin 
flip results in a move from $S_n$ to $S_{n+1}$ is 
$V^{-1}\sum_{a b}\sum_{\sigma\ne 1} C^\sigma_{a b}(n,i) p^{\sigma 1}_{a b}$, 
where the factor $V^{-1}$ represents the probability of choosing one 
particular spin.  Consequently, 
\begin{eqnarray}
\lefteqn{ 
{1\over V} 
\sum_{i=1}^{|S_n|} W(n,i) \sum_{ab} \sum_{\sigma\ne 1} 
 C^{\sigma}_{ab}(n,i) p_{ab}^{\sigma 1} = } \nonumber \\
 & & N_{\rm esc} +
 {1\over V}
 \sum_{i=1}^{|S_{n+1}|} W(n+1,i) 
 \sum_{ab} \sum_{\sigma\ne 1} C^{1}_{ab}(n+1,i) p_{ab}^{1 \sigma} 
\end{eqnarray}
Since each $S_n$ must be visited at 
least $N_{\rm esc}$
times, we can multiply the two terms involving sums by an identity and 
write the above equation as 
\begin{eqnarray}
\label{BIGONE}
\lefteqn{ 
{\sum_{j=1}^{|S_n|} W(n,j) \over \sum_{k=1}^{|S_n|} W(n,k)}
\sum_{ab} \sum_{\sigma\ne 1} p_{ab}^{\sigma 1} 
\sum_{i=1}^{|S_n|} W(n,i)  c^{\sigma}_{ab}(n,i)  = N_{\rm esc}  
} \nonumber \\
 & & + {\sum_{j=1}^{|S_{n+1}|} W(n+1,j) \over \sum_{k=1}^{|S_{n+1}|} W(n+1,k)}
\sum_{ab} \sum_{\sigma\ne 1}  p_{ab}^{1 \sigma} 
\sum_{i=1}^{|S_{n+1}|} W(n+1,i) c^{1}_{ab}(n+1,i) \ .
\end{eqnarray}
In Eq.~({\ref{BIGONE}}) one recognizes the average lumped class concentration 
\begin{equation} 
c^{\sigma}_{ab}(n)  = { \sum_{i=1}^{|S_n|} W(n,i)  c^{\sigma}_{ab}(n,i) 
                   \left /
                         \sum_{k=1}^{|S_n|} W(n,k) \right .} 
\end{equation}
which enters 
the definition of the global spin-flip rates, Eq.~(\ref{eq:gsrates}).  
We divide by 
$V N_{\rm esc}$ 
to express the sojourn times in units of MCSS.  
This gives 
\begin{equation}
 h(n)
\sum_{ab} \sum_{\sigma\ne 1} p_{ab}^{\sigma 1} c^{\sigma}_{ab}(n)  =
 V^{-1} +
 h(n+1)
\sum_{ab} \sum_{\sigma\ne 1}  p_{ab}^{1 \sigma} c^{1}_{ab}(n+1)  ,
\end{equation}
which leads directly to 
Eq.~({\ref{TAUETC}}). Thus, in the absence of forcing, our formula
for the lifetime becomes exact.

\vskip 2.0 true cm
\begin{center}
{\bf Figure Captions}
\end{center}

\smallskip
Fig.~1.  
The measured average lifetime, $\tau$, is shown as a function of 
the inverse forcing rate.  This figure illustrates that 
excellent values for $\tau$ are obtained even with 
moderately fast forcing.  The figure is for $10^2$--$10^3$ escapes from 
the metastable state.  
To achieve uniform accuracy, more escapes are needed at high forcing rates.  

\end{document}